\documentclass[final,3p,times]{elsarticle}

\usepackage{amssymb}
\usepackage{hyperref}
\usepackage{geometry}
\usepackage{amsmath}
\usepackage{xcolor}
\usepackage{natbib}
\usepackage{subcaption}

\hypersetup{colorlinks=true,
            linkcolor=blue,
            urlcolor=blue,
            linktoc=all,
            citecolor=blue}
            
\journal{Astronomy $\&$ Computing}
\begin{document}

\begin{frontmatter}


\title{Tracing correlations between galaxy properties across the Cosmic Web: An IllustrisTNG-based study}

\author[first]{Anindita Nandi}
\ead{anindita.nandi96@gmail.com}

\author[first]{Biswajit Pandey}
\ead{biswap@visva-bharati.ac.in}

\author[second]{Prakash Sarkar}
\ead{prakash.sarkar@gmail.com}

\affiliation[first]{organization={Department of Physics, Visva-Bharati University},
            city={Santiniketan},
            postcode={731235}, 
            state={West Bengal},
            country={India}}
\affiliation[second]{organization={Department of Physics, Kashi Sahu College},
            city={Seraikella},
            postcode={833219}, 
            state={West Bengal},
            country={India}}
\begin{abstract}
We explore the impact of cosmic web environments on galaxy properties such as $(u-r)\,$colour, stellar mass, star formation rate, and stellar metallicity, using a stellar mass-matched sample of simulated galaxies from the Illustris TNG simulation.
We use Normalized Mutual Information (NMI) to quantify correlations among galaxy properties and apply Student’s t-test to assess the statistical significance of their differences across cosmic web environments. In every case, the null hypothesis is rejected at $> 99.99 \%$ confidence, providing strong evidence that correlations among galaxy properties are strongly dependent on cosmic web environments.

\end{abstract}



\end{frontmatter}




\section{Introduction}
Galaxies are the primary building blocks of the observable universe. According to the $\Lambda$CDM cosmological model, galaxies form and evolve within dark matter halos. These halos emerge from primordial density fluctuations in the early universe and grow hierarchically through gravitational collapse and successive mergers. As halos develop, they accrete gas from  surroundings, which subsequently cools and condenses at their centers, eventually forming galaxies~\citep{rees77,silk77,white78,fall80}. 
Despite significant progress in cosmological research, disentangling the complex processes that govern galaxy evolution continues to be a major challenge. One of the most effective way of investigating their evolution is to analyze different physical properties of galaxies, which serve as direct diagnostics of their physical state and past history. These properties are shaped by both internal processes and external physical mechanisms.
Galaxies undergo secular evolution through various internal processes including stellar evolution, supernovae explosions and gas accretion onto supermassive black holes, leading to active galactic nuclei (AGN) activity. Feedback from both supernovae and AGN plays crucial role in regulating the inflow and outflow of gas, thereby influencing star formation activity of galaxies.
Gas inflow supplies the raw fuel for star formation. It can trigger bursts of star formation, influencing the mass, colour and morphology of galaxies. Moreover, the inflowing gas affects the chemical composition of galaxies by supplying low metallicity material from the surrounding intergalactic medium. 
In contrast, gas outflows can expel gas from galaxies, effectively reducing the available gas reservoir for star formation. In extreme cases, strong outflows can lead to the quenching of star formation in galaxies. Furthermore, two internal mechanisms, morphological quenching~\citep{martig09} and bar quenching~\citep{masters10} can also halt star formation in galaxies.
There are also several external mechanisms which play a pivotal role in regulating galaxy properties, arise from interactions between galaxies and their surrounding environment. The local environment of galaxies is generally characterized by local overdensity, which has been shown to correlate with various galaxy properties, such as morphology~\citep{dressler80}, colour~\citep{kauffmann04,baldry06} and star formation rate~\citep{grutzbauch11,koyama13,schaefer17,tanaka04}. Numerous studies using simulations~\cite{toomre72,mihos95,barnes96,tissera02,cox05} and observational data~\cite{larson78,barton99,alonso04,nikolic04,woods06,woods07}, find that tidal torques from galaxy interactions can induce starbursts and alter the colour and morphology of galaxies. In addition, the mass of the host dark matter halo influence the galaxy properties through different physical mechanisms such as mass quenching~\citep{birnboim03, dekel06, keres05, gabor10} and angular momentum quenching~\citep{peng20}. Other external processes including galaxy mergers~\citep{hopkins08}, starvation~\citep{larson80, somerville99, kawata08}, harassment~\citep{moore96,moore98} are also effective at impacting the star forming activity of galaxies.\par
However, the environment of a galaxy extends beyond its immediate surroundings. The three-dimensional distribution of galaxies, as revealed by different spectroscopic surveys~\citep{colless01,stoughton02} shows that galaxies are embedded 
within a vast interconnected  structure known as the `Cosmic Web'~\citep{gregory78,joeveer78a,joeveer78b,einasto80,zeldovich82,lapparent86,bond96,bharadwaj99,pandey05,aragoncalvo10,libeskind18,sarkar19}. The cosmic web is composed of four distinct parts, void, sheet, filamet and cluster. Voids are the large underdense regions which occupy most of the cosmic volume. Filaments are found at the  intersections of sheets or walls, while clusters form at the junctions of filamentary regions. Therefore, local environment does not give the full picture of the galaxy evolution. N-body simulation results indicate a flow of matter through the cosmic web, moving from voids to walls, walls to filaments and finally into clusters~\citep{aragoncalvo08,cautun14,ramachandra17,wang24}. Each of these components  provides a unique large-scale environment for galaxies.
Voids are pristine low density environments that foster distinct evolutionary pathways for galaxies~\citep{rodriguez25}, in contrast to those in other cosmic web environments. Galaxies in voids are generally late-type, having lower mass, higher star formation rate and younger stellar population~\citep{rojas05,hoyle05,ceccarelli08,ricciardelli14,rodriguez24}. Studies from hydrodynamical simulations ~\citep{tuominen21,espinosa21} indicate that 
filamets are rich in the warm-hot intergalactic medium (WHIM) gas, which accounts for more than $\sim 40-50\%$ baryonic matter. The galaxies located in different types of large-scale environments may have different gas accretion efficiency. Galaxies located near the centers of cosmic filaments and sheets often experience continuous inflow of cold gas, promoting active star formation and growth in stellar mass~\citep{chen17,pandey20,singh20,laigle21,das23a,das23b,hoosain24}. However, recent studies utilizing hydrodynamical simulations~\citep{dave19,nelson19} suggest that the shock heating in filaments can also inhibit star formation activity, particularly in satellite galaxies~\citep{bulichi24,hasan24}. Clusters are the high-density regions that form at the nodes, where filaments intersect. Galaxies located in clusters typically exhibit reduced star formation as a result of gas stripping due to the hot intracluster medium~\citep{treu03,poggianti06}. However,~\citep{donnan22}, using simulation data~\citep{nelson19} finds that galaxies which are in close proximity to cluster regions, show higher gas-phase metallicity compared to what is observed in filaments.\par
Numerous studies~\citep{zehavi11, yan13, alpaslan15, alam19} find that the assembly history of dark matter halos and differences in halo mass functions, have a greater impact in shaping galaxy properties compared to the cosmic web environments. Conversely, there are also many studies using data from different galaxy surveys e.g. Sloan Digital Sky Survey (SDSS)~\citep{pandey06,pandey08,paz08,jones10,scudder12,tempel13a,tempel13b,filho15,luparello15,pandey17,pandey20,kuutma17,chen17,chen19,lee18,kraljic20,bonjean20,winkel21}, Two-Degree Field Galaxy Redshift Survey (2dFGRS)~\citep{trujillo06}, Two Mass
Redshift Survey (2MRS)~\citep{erdogdu07},  Galaxy And
Mass Assembly~\citep{alpaslan16,kraljic18,bhambani23}, Cosmic Evolution Survey (COSMOS)~\citep{darvish14,laigle18} and hydrodynamical simulations~\citep{malavasi22,pandey24}, indicating the influence of large-scale environments on galaxy properties. 
Apart from their individual dependencies on large-scale environment, these galaxy properties such as stellar mass, star formation rate, colour, and metallicity exhibit strong correlations with each other. For instance, blue galaxies tend to be lower-mass systems having spiral morphologies and active star formation, whereas red galaxies are generally more massive and have mostly ceased their star formation. However, the nature of these correlations is not universal and appears to be contignent on the large-scale environment. Recently, \citep{nandi24} utilize data from SDSS and show that the correlations between different observable galaxy properties are sensitive to the cosmic web environments.
While~\citep{nandi24} is based on observational data, simulations serve as valuable tools for exploring correlations between galaxy properties, without being affected by observational limitations such as redshift-space distortions and measurement uncertainties.
Additionally, simulations typically yield a much larger number of galaxies in a single snapshot, which significantly improves the statistical robustness of correlation analysis across different cosmic environments. Importantly, it is also essential to assess whether the physical modeling implemented in simulations accurately reproduces the trend observed in real data. Comparing simulation results with observations serves as a critical test of the underlying physical prescriptions and helps refine our theoretical understanding of galaxy formation and evolution.\par
In this work, we use the state-of-the-art cosmological hydrodynamical simulation, IllustrisTNG~\citep{nelson19}. The IllustrisTNG simulation suite is one of the most widely used cosmological hydrodynamical simulations available for studying galaxy formation and evolution. It combines large cosmological volumes with high resolution and incorporates a comprehensive set of physical processes, including magnetohydrodynamics and scheme for black hole feedback, galactic winds, stellar evolution and gas chemical enrichment~\citep{pillepich18a,pillepich18b}. Utilizing data from IllustrisTNG, we analyze correlations between stellar mass, star formation rate, $(u-r)\,$ colour and metallicity. As the correlations between various galaxy properties are non-linear, we employ normalized mutual information (NMI)~\citep{strehl02} to quantify these correlations. The principal advantage of NMI is its sensitivity to non-linear relationships without requiring any assumptions about the functional form of the variables.\par
The structure of the paper is as follows. In \autoref{sec:data} we describe the simulation data used in our analysis. The methods employed for this study is outlined in \autoref{sec:method}. We present the results in \autoref{sec:results}, and finally \autoref{sec:conclusions} summarizes our conclusions.

\section{Data}
\label{sec:data}
IllustrisTNG (\textit{The Next Generation} Illustris; hereafter TNG)~\citep{pillepich18b,nelson17,marinacci17,naiman18,springel17}, 
an updated version of the original Illustris project, is a suit of cosmological simulations. TNG is based on the moving-mesh code AREPO~\citep{springel10},
with an improved galaxy formation model described by ~\citep{pillepich18a,weinberger17}. TNG adopts a $\Lambda$CDM cosmology with the following parameter 
values, $\Omega_m = 0.3089$, $\Omega_{\Lambda} = 0.6911$, $\Omega_b = 0.0486$, $n_s = 0.9667$, $\sigma_8 = 0.8159$ and $h=0.6774$ \citep{planck15}. 
There are three different simulation volumes available with $50$ Mpc, $100$ Mpc and $300$ Mpc side length, each with their lower-resolution and dark-matter only variations. In this analysis, we utilize data from the highest resolution realization of the largest volume TNG simulation, i.e. TNG300-1, with a baryon mass resolution $\sim 1.1 \times 10^7\,M_{\odot}$ and unit dark matter particle mass $\sim 5.9 \times 10^7\,M_{\odot}$. In the present analysis, we study the following four galaxy properties - $(u-r)\,$ colour, stellar mass, star formation rate (SFR) and metallicity.\par
Galaxies (subhalos) are identified using the SUBFIND algorithm~\citep{springel00,dolag08}. The galaxy colours are obtained from the catalog ``SDSS ugriz and UVJ
Photometry/Colors with Dust" prepared by~\cite{nelson17}, which accounts for the effect of dust attenuation. For the stellar mass of a galaxy, we consider the sum of the masses of all stellar particles within twice the stellar half-mass radius ($r_{stars,1/2}$)~\citep{pillepich18b}. The stellar half-mass radius is defined as the radius of a sphere that encloses half of the total stellar mass of a galaxy. The star formation rate is calculated by summing up the individual SFRs of all gas cells considering the same radial extent. Here we use stellar metallicity ($M_Z/M_{tot}$) of galaxies, which is defined as the mass weighted average metallicity of the stellar particles within $2 r_{stars,1/2}$. We select galaxies in the stellar mass range $10^9 \leq M_{\star} \leq 10^{12}$, to ensure a sample of well-resolved galaxies with at least one hundred stellar particles. The total number of galaxies in our final sample is $223442$.

\begin{figure}[htbp]
\centering
  \centering
  \includegraphics[width=0.5\linewidth]{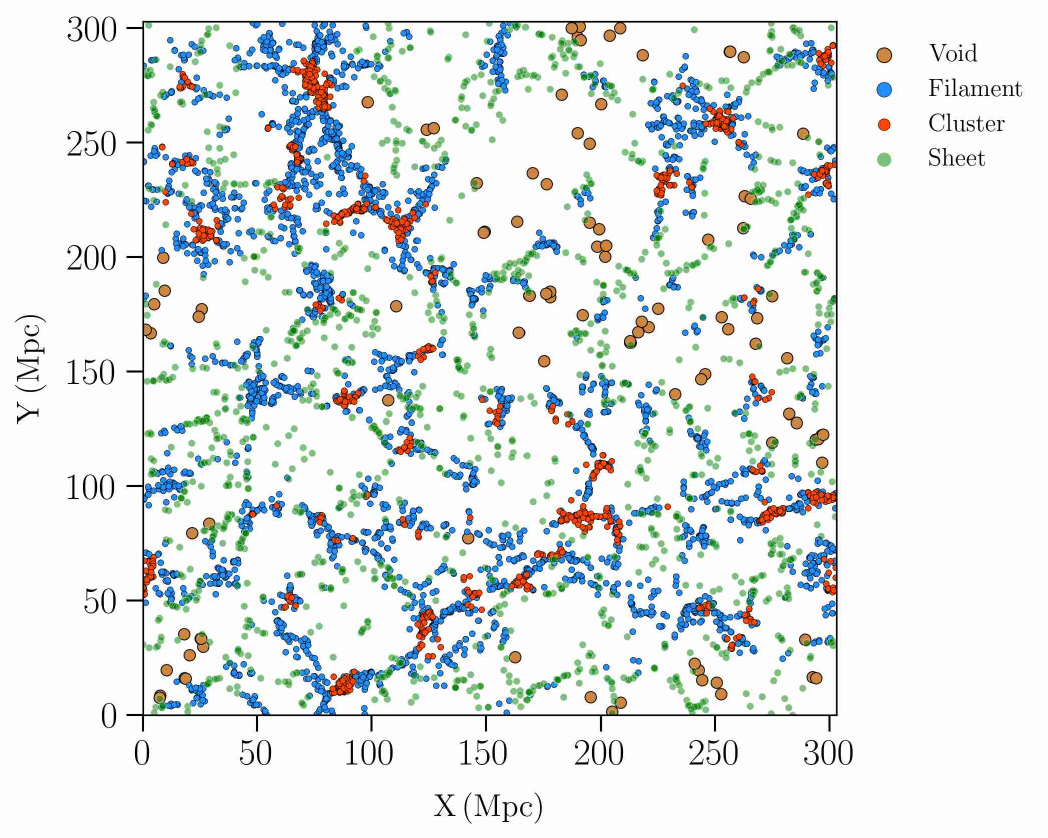}
  \label{fig:cosmic_web}
\caption{This figure shows a projected distribution galaxies in different cosmic web environments within a slice of thickness 10 Mpc in TNG300.}
\label{fig:cw_classify}
\end{figure}


\section{Methodology}
\label{sec:method}
\subsection{Identifying different components of the cosmic web}
The cosmic network of the galaxies is classified into its different parts, e.g. void, sheet, filament, and cluster following a 
Hessian-based approach~\citep{hahn07, romero09}. This method utilizes the Hessian of the gravitational potential, defined as,
\begin{equation}
    T_{ij} = \frac{\partial^2 \Phi}{\partial x_i \partial x_j}\,.
    \label{eq:detens}
\end{equation}
In \autoref{eq:detens}, $T_{ij}$ is the deformation tensor and $\Phi$ is the gravitational potential. The gravitational potential is 
obtained by solving the Poisson equation,
\begin{equation}
    \nabla^2 \Phi \equiv \delta \, ,
\end{equation}
where $\delta = \frac{\rho - \bar{\rho}}{\bar{\rho}}$ is the density contrast or overdensity.
We construct the overdensity field from the galaxy distribution on $(256)^3$ grids, using the Cloud-In-Cell (CIC) algorithm. 
The overdensity field is then transformed into the Fourier space and smoothed by applying an isotropic Gaussian filter having radius $4$ times
 the grid spacing,
which is close to the intergalactic separation of our sample $(\approx 4.79\, \rm  Mpc)$. This puts a limit on the capacity of describing
the cosmic web environments on scales below than the intergalactic separation. In this analysis, we are mainly concerned with quantifying 
the large-scale geometric environments of the cosmic web.\par
The gravitational potential associated with the fluctuations in the smoothed density field is determined by 
\begin{equation}
    \hat{\Phi} = \hat{\mathcal{G}}\hat{\rho}\,.
\end{equation}
Here $\hat{\mathcal{G}}$ is the Fourier transform of the Green's function of the Laplacian operator and $\hat{\rho}$ is the density in the Fourier space. We transform back the gravitational potential into the real space and perform numerical differentiation to obtain the Hessian of the gravitational potential. Next, we compute the eigenvalues $(\lambda_1, \lambda_2, \lambda_3)$ of the deformation tensor. 
The galaxies are classified on the basis of the signs of the eigenvalues $( \lambda_1 > \lambda_2 > \lambda_3)$ as follows.
\begin{enumerate}
    \item Void, when $\lambda_1,\lambda_2,\lambda_3 < 0$
    \item  Sheet, when $\lambda_1 >0, \lambda_2,\lambda_3 < 0$
    \item Filament, when $\lambda_1,\lambda_2 > 0, \lambda_3 < 0$
    \item Cluster, when $\lambda_1, \lambda_2, \lambda_3 > 0$
\end{enumerate}

The total number of galaxies identified in each cosmic web environment are provided in \autoref{tab:gals_in_web}.

\begin{table}[h!]
    \centering
    \begin{tabular}{c c}
    \hline
       Cosmic web environment  & Number of galaxies  \\
       \hline
       \hline
        Void & $2427$ \\
        \hline
        Sheet & $45057$ \\
        \hline
        Filament & $128991$ \\
        \hline
        Cluster & $76073$ \\
        \hline
    \end{tabular}
    \caption{This table represents the number of galaxies identified in different cosmic web environments.}
    \label{tab:gals_in_web}
\end{table}
\subsection{Stellar Mass Matching and Controlling for local density}
\begin{figure}[htbp]
\centering
\includegraphics[width=0.8\textwidth]{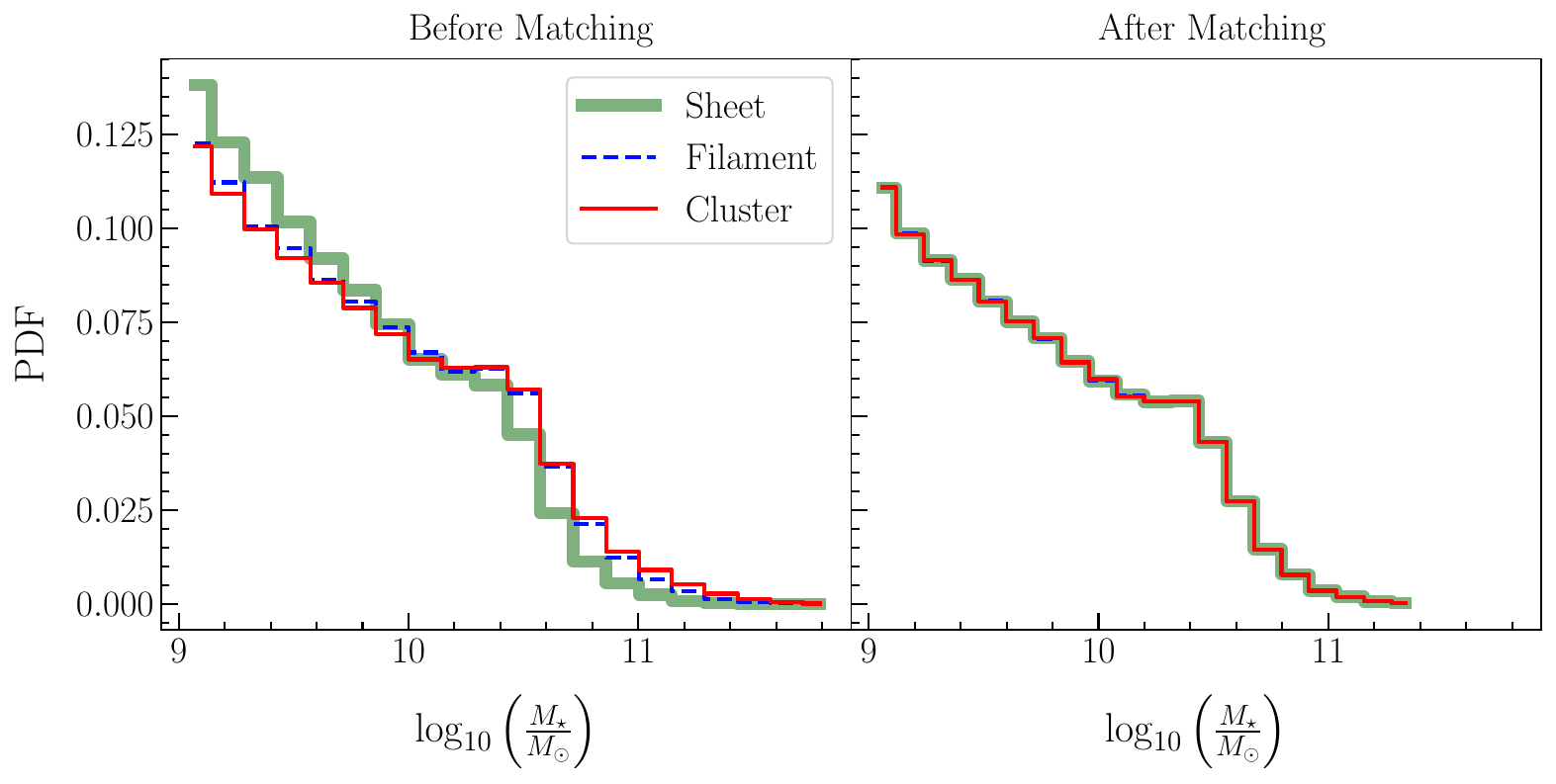}
\caption{The left panel and right of this figure shows the PDFs of $\log_{10}\left(\frac{M_{\star}}{M_{\odot}}\right)$, before and after matching the stellar masses of the galaxies, respectively.}
\label{fig:mass_matching}
\end{figure}

The stellar mass and local density play a crucial role in shaping galaxy properties. To ensure that any observed trends are primarily due to differences in the cosmic web environments, rather than being driven by variations in mass or density, we try to carefully control for both. We first restrict our analysis to a common range of local density across sheet, filament and cluster. We then perform stellar mass matching between galaxies in these three environments. We consider the sheet galaxies as our reference and perform the stellar mass matching such that the stellar mass difference between any two galaxies from different environments does not exceed 0.001 dex. We do not consider the void galaxies in our analysis due to their lower abundance. The distributions of $\log_{10}\left(\frac{M_{\star}}{M_{\odot}}\right)$ for the galaxies in sheet, filament and cluster are shown in \autoref{fig:mass_matching}. To ensure the robustness of our stellar-matching across different cosmic web environments, we perform the Kolmogorov-Smirnov (K-S) test to assess the statistical indistinguishability of the stellar mass distributions between each pair of environments. The results show that the null hypothesis can be accepted with a very high confidence level $(> 99.99 \%)$. 
As a consequence of the stellar mass matching process, each environment contains equal number ($37965$) of galaxies.

\subsection{Study of galaxy property correlation using Normalized Mutual Information (NMI)}
In information theory, mutual information (MI) is a measure of association between two random variables. MI can identify any kind of dependency,
including non-linear and non-monotonic correlations. Mutual information is related to the Shannon entropy~\citep{shannon48} of a random variable.
If $X$ is a discrete random variable having $n$ possible outcomes \{$X_i: i=1,\ldots n$\} and $P(X_i)$ is the probability corresponding to the 
$i^{th}$ outcome, then the Shannon entropy related to $X$ is given by,
\begin{equation}
    H(X) = -\sum_{i=1}^n P(X_i)\log P(X_i)\, ,
\end{equation}
where we have considered logarithm of base $10$.
Let $X$ and $Y$ represent two different galaxy properties and $P(X,Y)$ is the joint probability between them, then the join entropy corresponding to them is defined as,
\begin{equation}
    H(X,Y) = -\sum_{i=1}^{n_1}\sum_{j=1}^{n_2}P(X_i,Y_j)\log P(X_i,Y_j) \, .
\end{equation}
We consider $n_1, n_2 = 30$ in this analysis.\\
The mutual information is expressed as,
\begin{equation}
    I(X;Y) = H(X) + H(Y) - H(X,Y) \, .
\end{equation}
\par
We quantify the correlations between different galaxy properties using normalized mutual information (NMI)~\citep{strehl02}, which is the normalized form of mutual information,
\begin{equation}
    NMI(X;Y) = \frac{I(X;Y)}{\sqrt{H(X)H(Y)}}
\end{equation}

\section{Results}
\label{sec:results}

\subsection{Exploring correlations between galaxy properties in different parts of cosmic web}

\begin{figure}[htbp]
\centering
\includegraphics[width=1.0\textwidth]{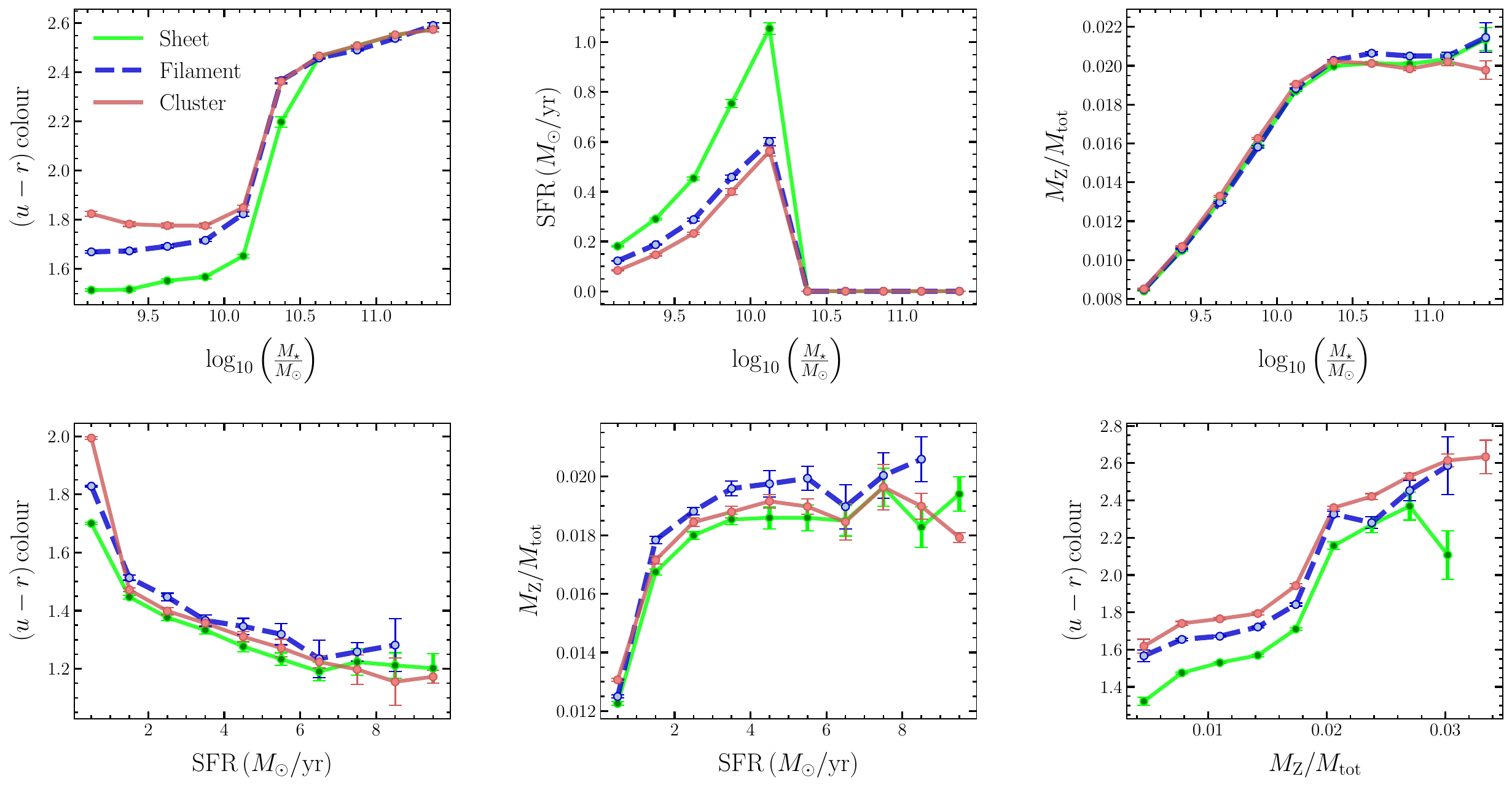}
\caption{The different panels in this figure illustrate the relationships between various pairs of galaxy properties. The y-axis in each panel represents the median value of a galaxy property, computed with bins of another galaxy property shown along the x axis. Only those bins with at least 10 galaxies are considered for median calculation. The $1\sigma$ errors on each data points are estimated from 50 bootstrap samples drawn from the original dataset. }
\label{fig:medians}
\end{figure}
We investigate the correlations among $(u-r)\,$ colour, stellar mass, star formation rate, and stellar metallicity of galaxies in different cosmic web environments. In \autoref{fig:medians}, we illustrate these correlations by showing the median of one property as a function of the other. Specifically, we bin one galaxy property and then compute the median of the other galaxy property in each bin, considering only those bins that contain at least $10$ galaxies.\par
In the top left panel of \autoref{fig:medians}, the median $(u-r)\,$ colour is shown as a function of stellar mass for three different cosmic environments - sheet, filament and cluster. Overall, the median $(u-r)\,$colour increases with stellar mass across all environments. While the general trend is similar, the rate of increment depends on the stellar mass range. There is a sharp rise around stellar mass of $\sim 10^{10.2}\,M_{\odot}$ in all environments. Below this mass threshold, the median colour changes very slowly with stellar mass, regardless of the environment. However, there is a systematic vertical shift in median colour, with galaxies in clusters showing highest values, followed by filaments and sheets. Above this mass threshold, the median $(u-r)\,$colour increases rapidly with stellar mass before eventually saturating around  $\sim 10^{10.6}\,M_{\odot}$.
The median colour at lower masses rises systematically across environments, being strongest in clusters, intermediate in filaments, and weakest in sheets, a pattern likely driven by local density variations. Even though the analysis is restricted to a common density range, the underlying distributions of local densities vary among the different cosmic web environments. Lower stellar mass galaxies, which are mostly satellites, are more strongly affected by environment, whereas higher mass galaxies show no significant differences.This highlights that the colour transformation is dominated by environment-driven mechanisms for lower mass galaxies and by mass-driven processes for massive galaxies.
In the top middle panel of \autoref{fig:medians}, we present the median star formation rate as a function of stellar mass for galaxies in different environments. The median SFR gradually increases, reaching a peak value around $\sim 10^{10}\,M_{\odot}$. Beyond this peak, the median SFR steadily declines and approaches zero at stellar masses above $10^{10.5}\,M_{\odot}$. This behavior suggests a transition from active star-forming galaxies to more quiescent systems as stellar mass increases, irrespective of the cosmic environment. Below $\sim 10^{10}\,M_{\odot}$, although the SFR increases sharply with stellar mass across all environments, galaxies in sheets exhibit higher median SFRs, compared to their counterparts in filament and cluster.  
The right panel in the top row depicts the metallicity as a function of stellar mass for three different environments. The stellar metallicity of the simulated galaxies, which are measured here as the mass-weighted metal fractions calculated from stellar particles, shows a tight correlation with stellar mass up to $\sim 10^{10.4}\,M_{\odot}$, after which it flattens. Unlike $(u-r)\,$ colour and star formation rate, the variation of metallicity with stellar mass exhibits no such environmental dependence for lower mass galaxies.\par
The bottom left panel represents the median $(u-r)\,$ colour as a function of star formation rate. As expected, the colour values decrease with increasing star formation rate, indicating that galaxies with higher star formation rates tend to be bluer. Also we can see from this figure that galaxies in the lowest SFR bin, galaxies in clusters have relatively higher values of $(u-r)\,$ colour, followed by filaments and sheets.
This environmental trend is likely driven by the mass distribution of quiescent galaxies. The upper middle panel shows, the lowest SFR bin contains both low mass ($M_{\star} < 10^{9.2}$) and high mass ($M_{\star} > 10^{10.5}$) galaxies in all environments. The upper left panel reveals that clusters have an abundance of low mass redder galaxies. This population may be responsible for the elevated median colour of low-SFR galaxies in that environment.
The bottom middle panel presents the median metallicity as a function of star formation rate. Metallicity varies in a non-monotonic way with the star formation rate. Galaxies with higher star formation rates have higher metallicity compared to low star-forming galaxies. The galaxies in sheets are more metal rich than those in filaments and clusters in the  SFR range $2-6\,M_{\odot}/yr$.
We show the median $(u-r)\,$colour as a function of stellar metallicity in the bottom right panel. Consistent with the trends in the upper left and right panels, where $(u-r)\,$ colour and metallicity increase with stellar mass, we observe a positive correlation between colour and metallicity. This probably arises because massive galaxies which typically exhibit redder colour due to their lower star formation activity, also retain higher metallicity through their deeper potential wells. Conversely, Lower metallicities in low-mass galaxies arise from their gas reach, blue, star-forming nature and the fact that their shallow gravitational potentials make them prone to losing metals through supernova-driven outflows.

\subsection{Characterizing the correlations using normalized mutual information}
From \autoref{fig:medians}, it is evident that the correlations between various galaxy properties are not strictly linear. Therefore, to quantify such non-linear relationships we employ the NMI. NMI requires the computation of the probability distribution functions (PDFs) for each individual galaxy property and the joint (2D) PDFs for each pair of properties.

\begin{figure}[htbp!]
\centering
\includegraphics[width=.8\textwidth]{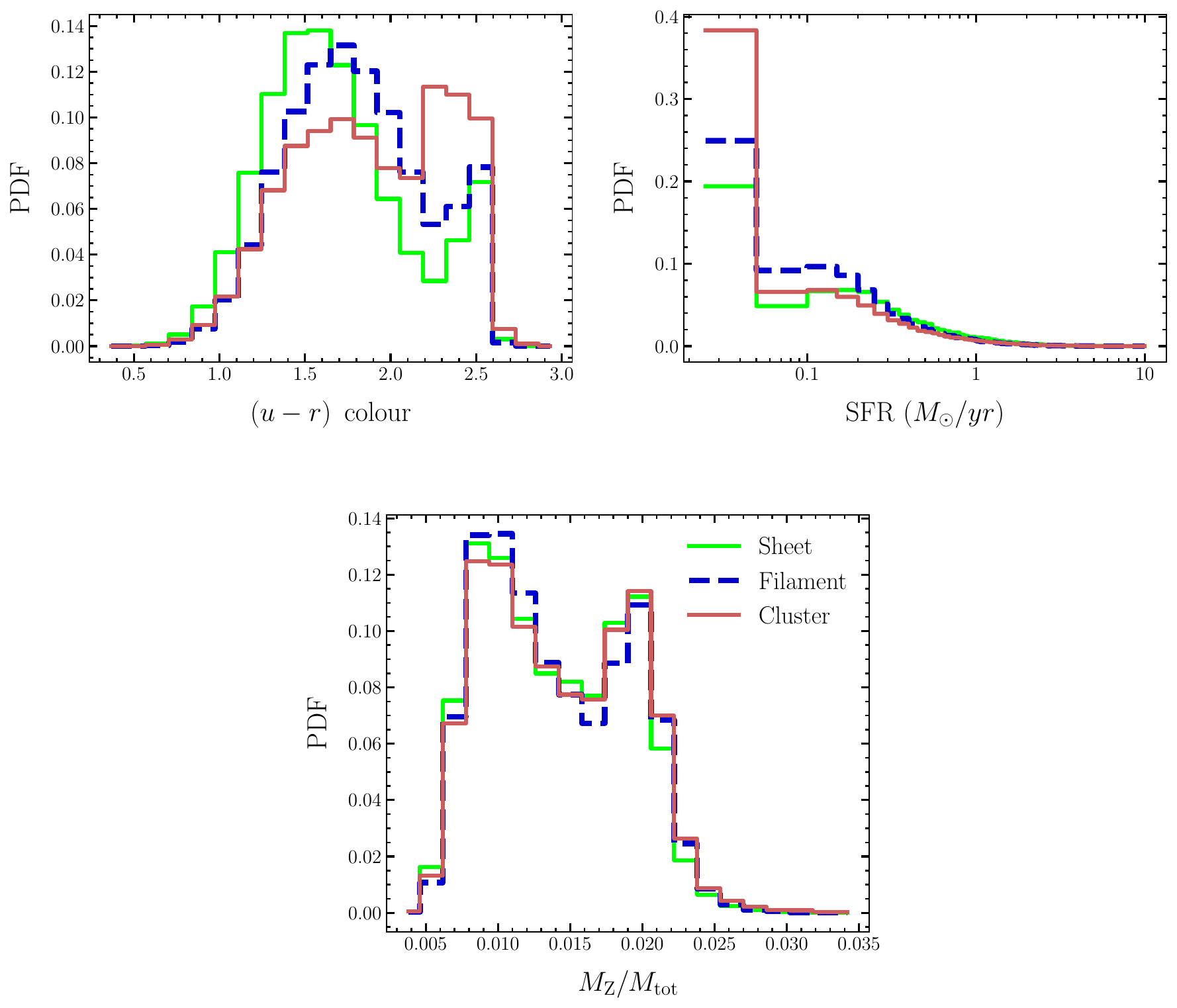}
\caption{The different panels of this figure show the probability
  distribution functions (PDF) of $\rm (u-r)$ colour, $\rm SFR$, and
  metallicity. The PDFs in sheets, filaments and clusters are shown
  together in each panel for comparison.}
\label{fig:pdfs}
\end{figure}

\begin{figure}[htbp!]
    \centering
        \includegraphics[width=1.\textwidth]{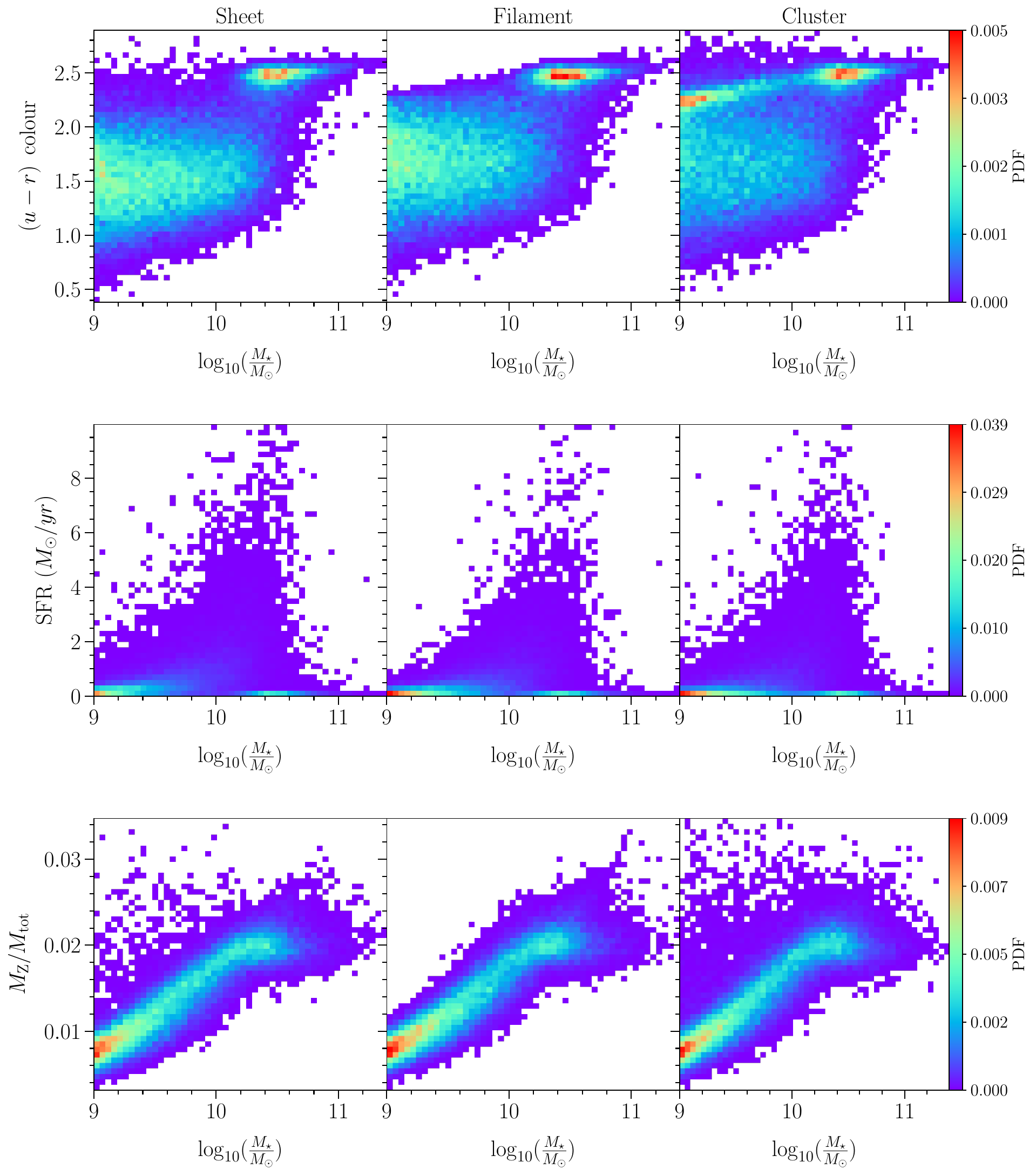}
        \caption{This figure illustrates the joint probability distributions of $(u-r)\,$ colour, star formation rate and stellar metallicity with stellar mass of galaxies, in three different types of cosmic web environments.}
    \label{fig:2dpdf_set1}
\end{figure}

\begin{figure}[htbp!]
    \centering
        \includegraphics[width=1.\textwidth]{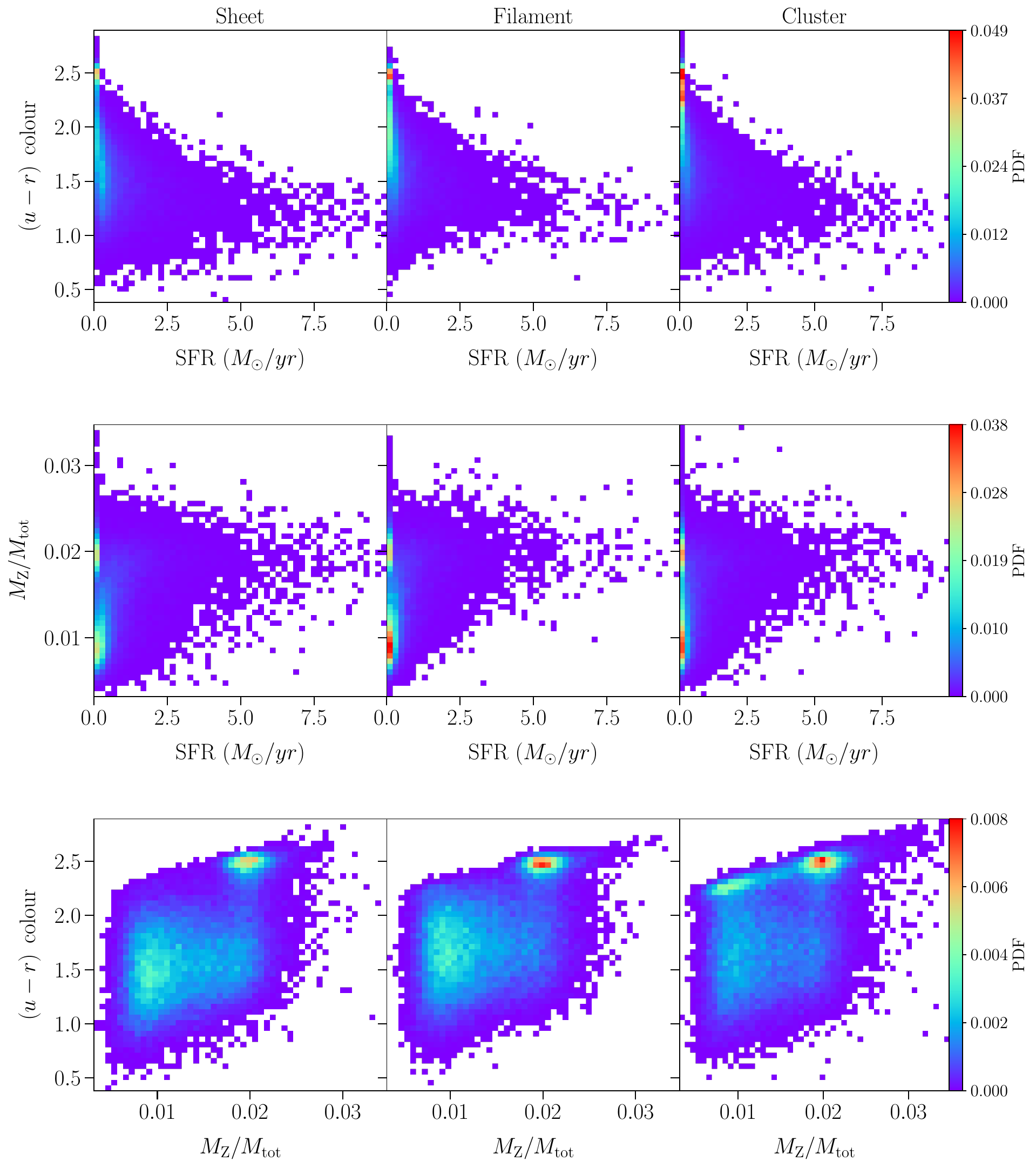}
        \caption{Same as \autoref{fig:2dpdf_set1}, but for different sets of galaxy properties including $(u-r)\,$ colour, star formation rate and metallicity.}
    \label{fig:2dpdf_set2}
\end{figure}

\begin{figure}[htbp!]
    \centering
    \includegraphics[width=0.9\textwidth]{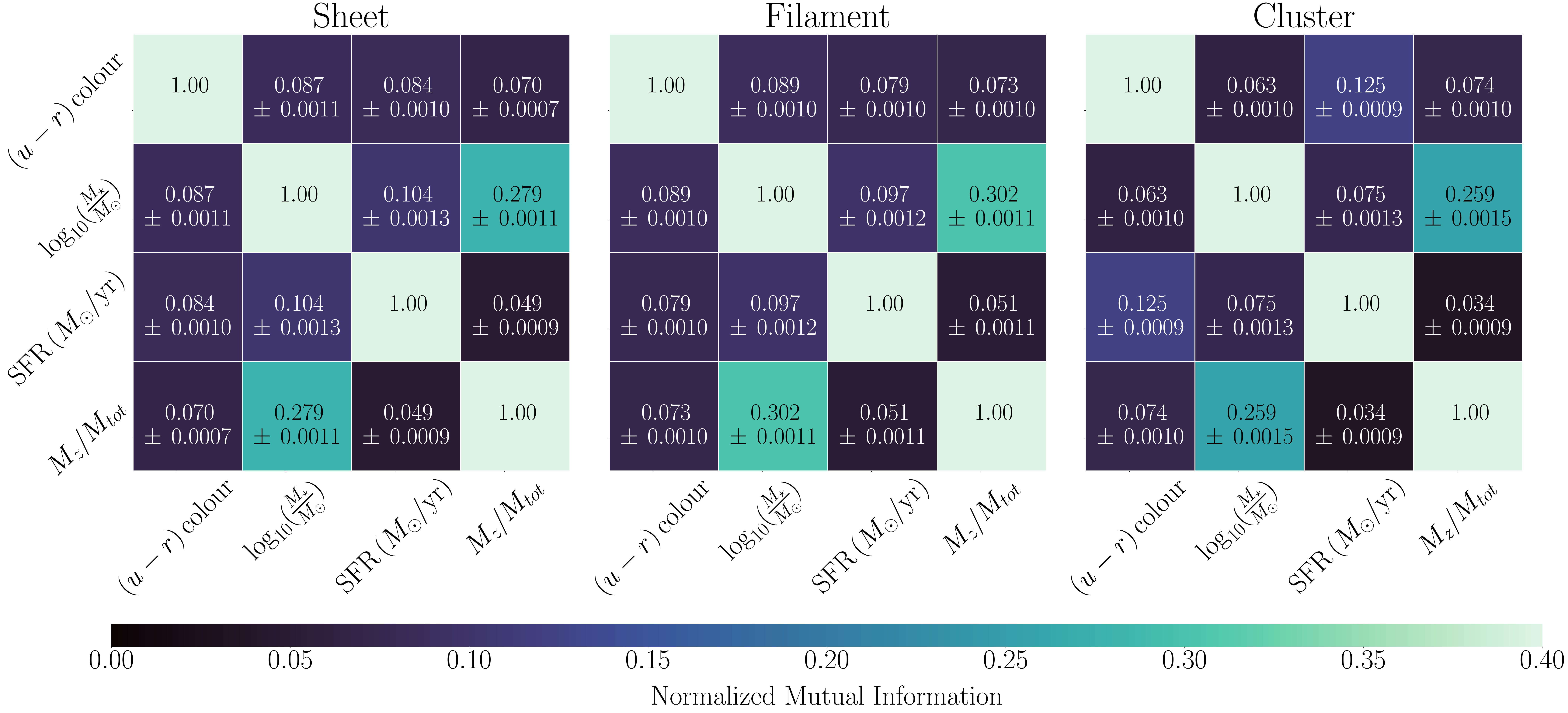}
    \caption{This figure represents the Normalized Mutual Information (NMI) for each pair of galaxy properties in three different cosmic web environments. The $1\sigma$ errorbars corresponding to these measurements are estimated utilizing 50 jack-knife samples to compute the errorbars.}
    \label{fig:nmi}
\end{figure}

\autoref{fig:pdfs} presents the probability distribution functions (PDFs) of $(u-r)\,$ colour, star formation rate (SFR), and metallicity across different galaxy environments. The top left panel shows the PDF of $(u-r)\,$ colour, where sheets dominate at lower values, followed by filaments, and then clusters. This trend persists up to a $(u-r)\,$ colour of 2. However, for colour values greater than 2, clusters dominate, followed by filaments and sheets. This distribution suggests that galaxies in less dense environments, like sheets and filaments, are typically bluer, indicating younger stellar populations and ongoing star formation. In contrast, galaxies in clusters tend to be redder, especially for (u-r) values greater than 2, implying older, more evolved populations with reduced star formation activity. The top right panel shows the PDF of SFR for different structural environments. At SFR values below 0.05, clusters dominate, followed by filaments and sheets. Filaments lead at SFR values between 0.05 and 0.1, followed by clusters and sheets, while sheets dominate at SFR values greater than 0.3 and less than 1, with filaments in second place. The number of galaxies drops significantly for SFR values greater than 1, across all environments. The distribution of SFR indicates that galaxies in clusters experience suppressed star formation, while those in filaments and sheets sustain moderate star formation rates. The lower panel shows the probability distribution function (PDF) of metallicity for the three different structural environments. The three environments exhibit a similar pattern, with sheets slightly dominating, followed by filaments and then clusters, up to a metallicity of 0.022. Beyond this point, clusters dominate, followed by filaments and then sheets. This pattern suggests that galaxies in less dense environments such as sheets and filaments exhibit less chemically enriched galaxies while high-density environments such as clusters show an accumulation of heavy elements. \par
In \autoref{fig:2dpdf_set1}, we represent the joint probability distribution functions (2D PDFs) for the combinations between different galaxy properties including $(u-r)\,$ colour, SFR, metallicity and $\log(\frac{M_{\star}}{M_{\odot}})$. The top panel shows the 2D PDF between $(u-r)\,$ colour and stellar mass in sheet, filament and cluster. There is a clear indication of bimodality in the colour-stellar mass plane in all these three environments, with a variation in the maximum populated regions with the web environments. A notable common feature among these environments is that, in the stellar mass range of ($10^{10.2}$ - $10^{10.8}$) and the colour range of ($2.35$-$2.55$), a higher concentration of galaxies exists.
Additionally, at lower stellar masses around ($10^9$ - $10^{9.4}$), a small overdense region can be identified in cluster environment, and there is a connecting bridge-like region between the higher mass-higher colour group and lower mass-lower colour group. Also, if we consider a specific mass bin at lower mass end, we find a broad dispersion in their $(u-r)\,$ colour values. This diversity in colour diminishes as we move toward higher stellar masses. This transition of galaxies from star-forming to quiescent state is most likely due to the AGN feedback as implemented in TNG model~\citep{zinger20, donnari21}.
In the middle panel, the distribution of galaxies in the star formation rate-stellar mass plane reveals quiescent galaxies bifurcating into two subpopulations, one at lower masses and the other at higher stellar masses.
Although the variation in colour values is maximum corresponding to the lowest mass bin, we can not find any such high dispersion in star formation rates of galaxies. The peak in SFR variation occurs at stellar mass $\sim 10^{10.4}\,M_{\odot}$.
The bottom panel in \autoref{fig:2dpdf_set1} presents the distribution of galaxies in the stellar mass-metallicity plane across three different environments. While the joint distributions are similar across these environments, the scatter in this plane is smallest for the filament-type environment, in contrast to the broader distributions observed in sheets and clusters. The combination of the middle and bottom panels of \autoref{fig:2dpdf_set1} suggests that the lower-mass quiescent systems are generally metal-poor, whereas their higher mass counterparts are predominantly metal-rich.

In \autoref{fig:2dpdf_set2}, we show the joint distribution of galaxies in three different parameter spaces, the $(u-r)\,$ colour - star formation rate plane (top panel), stellar metallicity-star formation rate plane (middle panel) and $(u-r)\,$ colour-metallicity plane (bottom panel). In the top panel, we find that a population of low star-forming, redder galaxies is present in all environments. However, its dominance in clusters underscores the importance of environment-driven quenching mechanisms.  The bottom panel indicates that the stellar metallicity strongly correlates with galaxy colour. Specifically, galaxies with low metallicity are typically bluer, while those with higher metallicity appear redder. Within the cluster environment, we further identify a transitional population connecting these two regions as seen in the bottom panel of \autoref{fig:2dpdf_set2}. This trend resembles the distribution of galaxies in colour-stellar mass plane in the cluster environment, as shown in the top right panel in \autoref{fig:2dpdf_set1}. 
The middle and bottom panels of \autoref{fig:2dpdf_set2} also reveal the presence of both low metallicity and high metallicity quenched population.


Figure \ref{fig:nmi} illustrates the NMI between different galaxy properties across various cosmic environments. The left panel shows the NMI for sheets, followed by filaments in the middle panel and clusters in the right panel. The NMI values are displayed along with their corresponding $1\sigma$ errors. These errors were estimated using 50 Jackknife sub-samples of the original dataset. 
All pairs of galaxy properties exhibit a non-zero NMI. In all environments, almost all NMI values remain below 0.1, with the notable exception of the mass-metallicity relation, which shows a strong correlation as evident in the top right panel of \autoref{fig:medians} and the bottom panel of \autoref{fig:2dpdf_set1}. The NMI value for mass-metallicity relation is highest for filament, followed by sheet and cluster. This trend is also supported by the scatter between these two properties across three environments, as shown in \autoref{fig:2dpdf_set1}.
Among all galaxy property pairs, the correlation between star formation rate and metallicity is the weakest, which is minimum $(\sim 0.03)$ for cluster environment. We also observe that the NMI value quantifying the SFR-metallicity correlation is maximum in filament, followed by sheet and cluster.
To assess the statistical significance of the differences in the mean NMI values between various galaxy property pairs across
different cosmic web environments, we employ the Student’s t-test. For this purpose, we generate 50 jakc-knife samples for each environment, where each sample is created by randomly selecting 50\% of total galaxies in that environment. The mean NMI is then computed from these jack-knife realizations for every pair of environments. The $t$-score for each pair of properties is given by, 
\begin{equation}
t = \frac{\mu_1 - \mu_2}{\sigma_s\sqrt{\frac{1}{N_A} + \frac{1}{N_B}}}.
\end{equation}
Here $\mu_1$ and $\mu_2$ are mean values, $\sigma_s = \sqrt{\frac{(N_A-1)\sigma_1^2 + (N_B-1)\sigma_2^2}{N_A + N_B - 2}}$, is the standard error and $N_A$, $N_B$ are the total number of samples in the two environments being compared. We test the null hypothesis that, for a given pair of galaxy properties the mean NMI values in two separate environments are not significantly different. We perform a two-tailed t-test with a significance level of $\alpha = 0.001$ corresponding to a $99.9$ per cent confidence level. The degrees of freedom in this case is $(N_A + N_B - 2)$. If the $p$-value is less than $\alpha$, the null hypothesis is rejected, indicating statistically significant difference between two environments.
 The $t$-scores and $p$-values for each relation in each pair of geometric environments, as displayed in  \autoref{tab:nmi_ttest} indicate that the null hypothesis can be rejected for all relations in this case.\\
\begin{table}[htbp!]
\centering
\begin{tabular}{|c|c|c|c|c|c|c|}
\hline
{\rule{0pt}{3ex} Relations} & \multicolumn{2}{|c|}{ Sheet - Filament } & \multicolumn{2}{c|}{Filament - Cluster} & \multicolumn{2}{c|}{ Sheet - Cluster } \\
\cline{2-7}
& $t$ score & $p$ value & $t$ score & $p$ value & $t$ score & $p$ value \\
\hline\hline
\rule{0pt}{3ex} colour-stellar mass & $-6.16$ & $1.60 \times 10^{-8}$ & $125.80$ & $3.77 \times 10^{-110}$ & $114.29$ & $4.30 \times 10^{-106}$ \\
\hline
\rule{0pt}{3ex} colour-SFR & $27.80$ & $2.80 \times 10^{-48}$ & $-237.19$ & $4.78 \times 10^{-137}$ & $-210.40$ & $5.90 \times 10^{-132}$ \\
\hline
\rule{0pt}{3ex} colour-metallicity & $-16.19$ & $1.91 \times 10^{-29}$ & $-7.23$ & $1.06 \times 10^{-10}$ & $-24.54$ & $1.23 \times 10^{-43}$ \\
\hline
\rule{0pt}{3ex} stellar mass-SFR & $30.79$ & $3.42 \times 10^{-52}$ & $88.33$ & $3.13 \times 10^{-95}$ & $114.28$ & $4.32 \times 10^{-106}$ \\
\hline
\rule{0pt}{3ex} stellar mass-metallicity & $-104.88$ & $1.82 \times 10^{-102}$ & $164.48$ & $1.66 \times 10^{-121}$ & $74.89$ & $2.62 \times 10^{-88}$ \\
\hline
\rule{0pt}{3ex} SFR-metallicity & $-11.46$ & $8.73 \times 10^{-20}$ & $88.05$ & $4.25 \times 10^{-95}$ & $84.69$ & $1.84 \times 10^{-93}$ \\
\hline
\end{tabular}
\caption{This table shows the $t$-scores and $p$-values calculated using Student’s t-test after comparing the normalized mutual information between
different pairs of galaxy properties in two separate cosmic web environments. We have used a two-tailed t-test for our analysis. The degrees of
freedom in this test is $98$.}
\label{tab:nmi_ttest}
\end{table}

\section{Conclusions}
\label{sec:conclusions}
In this study, we examine the correlations between some key galaxy properties, such as $(u-r)\,$ colour, stellar mass, star formation rate and stellar metallicity in different cosmic web environments using redshift $z = 0$ snapshot of TNG300-1, the largest volume run of Illustris TNG simulation.
We identify the various cosmic web environments using the eigenvalues of the deformation tensor, estimated from the smoothed density field constructed from the discrete galaxy distribution. Before doing the analysis, we matched the stellar masses of the galaxies between sheet, filament and cluster and considered a shared density range to isolate the effects of cosmic web from stellar mass and local density.
We find that the galaxy properties e.g. colour, stellar mass, star formation rate and stellar metallicity exhibit strong correlations, which are modulated by the large-scale cosmic web environments hosting these galaxies. To quantify these non-linear and non-monotonic relationships, we employ the Normalized Mutual Information (NMI). Although the NMI values are small due to the large scatter among galaxy properties, the differences in NMI values across the three environments are statistically significant, suggesting that the correlations between galaxy properties are influenced by the large-scale environments. The observed differences in property correlations across environments can be linked to the varying efficiency of physical processes such as gas accretion, feedback processes and galaxy interactions, that are further linked to the geometry of the cosmic web. Our finding is consistent with previous studies using IllustrisTNG data~\citep{malavasi22, gala23, montero24, yu25}.
The analysis of individual and joint probability distributions of galaxy properties shows that $(u-r)\,$ colour, stellar mass and metallicity exhibit bimodal distributions, whereas the star formation rate is unimodal. \par
Comparing simulation results with observations is essential to test our understanding of the physical processes driving galaxy formation and evolution. We therefore compare our findings with \citep{nandi24}, where correlations between galaxy properties were analyzed using observational data, albeit with a different sample size and redshift range compared to the simulated galaxy data used here. The variation of median $(u-r)\,$ colour with stellar mass (see Figure 4 in \citep{nandi24}) is similar to what we observe in the top left panel of \autoref{fig:medians}.
The nature of distribution of $(u-r)\,$ colour, as shown in the top left panel of \autoref{fig:pdfs} also resembles the colour distribution of the observed galaxy sample presented in Figure 6 in \citep{nandi24}.
Additionally, the rate of change of specific star formation rate (sSFR) with stellar mass for galaxies with stellar mass $> 10^{10}\,M_{\odot}$, as shown in the top right panel in Figure 4 in \citep{nandi24}, is consistent with the trend seen in the top middle panel of \autoref{fig:medians}. 
However, the correlation between mass-metallicity is found to be very weak in \citep{nandi24}, contrary to what we find in the TNG data this case.
Our results align with several other observational and simulation-based studies, which underscore the impact of cosmic web environments on galaxy evolution (see~\citep{pandey24,nandi25a,nandi25b}, and references therein). 
For instance, in dense cluster regions, galaxies experience intense interactions and efficient quenching, leading to the earliest build up of the red sequence~\citep{pandey24}. Filaments act as highways, guiding galaxies before they enter clusters~\citep{rost21,kuchner22}.~\citep{aragoncalvo19} posits that galaxies gradually lose their connection to filamentary network, which limits fuel for star formation, driving quenching through starvation. 
Void galaxies, residing in the most underdense regions are predominantly late-type, gas rich and star-forming. 
Analysis using simulation data~\citep{bulichi24,yu25} suggest that satellite galaxies are more prone to environmental effects compared to central galaxies. While stellar mass~\citep{alpaslan15} and local density~\citep{okane24} are primary drivers of galaxy evolution, comparisons across cosmic web environments reveal additional modulations in colour, star formation rates, metallicity, morphology, suggesting that cosmic web exerts an influence beyond local density alone. Although different parts of the cosmic web inherently traces different density regimes, classifying galaxies solely on density risks erasing environmental signatures, since density distributions in different cosmic web types overlap substantially.
In summary, our analysis of TNG300-1 simulation data, demonstrates that the association between galaxy properties are indeed influenced by their large-scale cosmic web environment. However, we should acknowledge that since this work is based on IllustrisTNG data, our findings are inherently tied to a specific subgrid physics, neumerical methods and resolution of the adopted simulation. Also, our study is limited to the $z=0$ snapshot and restricted set of galaxy properties. A natural extension would be to trace these correlations across redshifts to examine how the role of cosmic web evolves over cosmic time. Incorporating additional properties such as galaxy morphology, gas content, black hole activity would provide a more complete picture. Moreover, the method that we have used to classify cosmic web environments involves subjective choices such as the smoothing scale, which can influence galaxy assignments and introduce minor variations in the reported correlations. Future comparisons with upcoming large galaxy surveys will be crucial for testing these predictions and further constraining the interplay between galaxies and their large-scale environments.\\


\section*{Acknowledgments}
AN acknowledges the financial support from the Department of Science
and Technology (DST), Government of India through an INSPIRE
fellowship. BP would like to acknowledge financial support from the
SERB, DST, Government of India through the project CRG/2019/001110. BP
would also like to acknowledge IUCAA, Pune, for providing support
through the associateship programme.
\appendix



\bibliographystyle{elsarticle-harv} 
\bibliography{ref}






\end{document}